\documentclass[aip,cha,reprint,numerical,onecolumn]{revtex4-1}
\usepackage{amsmath}
\usepackage{amssymb}
\usepackage{graphicx}
\usepackage{dcolumn}
\usepackage{bm}
\usepackage{natbib}		
\usepackage{color}

\newcommand {\e} {\varepsilon}
\newcommand {\vp} {\varphi}
\newcommand {\w} {\omega}

\begin{document}

\title{Nonlinear phase coupling functions: a numerical study}

\author{M. Rosenblum$^{1,2}$ and A. Pikovsky$^{1,2}$\\
$^{1}$Institute of Physics and Astronomy, University of Potsdam, 
Karl-Liebknecht-Str. 24/25, 14476 Potsdam-Golm, Germany\\
$^{2}$Control Theory Department, Institute of Information Technologies,
Mathematics and Mechanics, Lobachevsky University Nizhny Novgorod, Russia}

\keywords{Phase approximation, coupling function, phase response curve}


\begin{abstract}
Phase reduction is a general tool widely used to describe forced and interacting
self-sustained oscillators. Here we explore the phase coupling functions beyond the
usual first-order approximation in the strength of the force. Taking
the periodically forced Stuart-Landau oscillator as the paradigmatic
model, we determine and numerically analyse the coupling functions up to the fourth order in the
force strength. We show that the found nonlinear phase coupling functions
can be used for predicting synchronization regions of the forced oscillator.
\end{abstract}
\maketitle

\section{Introduction: Phase description of forced and coupled oscillators}
Models of coupled and forced self-sustained oscillators describe a variety of 
natural and social phenomena
and effects in man-made devices, ranging from synchronization of pendulum clocks, 
organ pipes, and electronic circuits to emergence of collective motion in populations 
of spin-torque or nanomechanical oscillators, neurons, yeast cells, pedestrians on footbridges, 
and synthetic genetic oscillators 
\cite{Winfree-80,Kuramoto-84,Blekhman-88,Hoppensteadt-Izhikevich-97,Pikovsky-Rosenblum-Kurths-01,%
  Strogatz-03, Balanov_et_al:2009,Breakspear-Heitmann-Daffertshofer-10,%
  Richard-Bakker-Teusink-Van-Dam-Westerhoff-96,Strogatz_et_al-05,Shim-Imboden-Mohanty-07,Mondragon-Palomino1315}. 
Probably the most important and frequently used theoretical tool for the analysis of
forced and interacting self-sustained units is the phase reduction method
\cite{Winfree-80,Kuramoto-84,Hoppensteadt-Izhikevich-97,Pikovsky-Rosenblum-Kurths-01,%
Ermentrout-Terman-10,Monga_Wilson-Matchen-Moehlis-18}.
This approach assumes that the force or the coupling is so weak that it does not essentiually
influence the amplitudes of the oscillators, but only their phases. 
The mathematical basis behind 
this assumption is the correspondence between the phase variable of an autonomous system 
and the neutrally stable direction (with zero Lyapunov exponent) along the limit cycle, 
while the amplitudes correspond to stable 
transversal directions, quantified by negative Lyapunov exponents. 
Hence, the effects of weak forcing can be described solely by an equation for 
the phases, while the amplitudes are enslaved. 

The theory of phase reduction in the first order in the strength
of the force is well established, 
see~\cite{Monga_Wilson-Matchen-Moehlis-18,Wilson-Ermentrout-18}
for recent reviews. In such an approximate description, 
the corresponding term in the 
phase dynamics equations, 
called the coupling function, scales linearly with the forcing/interaction strength. 
On the other hand, if the variations of the 
amplitudes due to the forcing and/or interaction 
cannot be neglected, but still the dynamics in the state space is confined to the 
surface of a smooth torus, the dynamical description in terms of the
phases is nevertheless possible. One cannot, however, expect the first order
perturbation theory to be valid for strong forcing, rather nonlinear effects should be visible. 
Although the derivation of the coupling functions
with account of relatively large deviations of the state space trajectory from the limit cycle
of an unperturbed system remains a theoretical challenge 
(see,
 e.g. \cite{Monga_Wilson-Matchen-Moehlis-18,Wilson-Ermentrout-18}), 
such nonlinear coupling functions can be estimated 
numerically, as have been demonstrated in our recent short 
communication~\cite{Rosenblum-Pikovsky-19}. Nonlinear coupling 
function depends nontrivially on the coupling strength, and, in contradistinction
to the linear coupling function, also depends
 on the frequency of forcing. 
A numerical exploration of these dependencies is the main
purpose of this paper.

\section{Phase dynamics models}
\label{sec2}
First, we briefly summarize the main results of the first-order phase approximation 
theory. Consider an autonomous self-sustained oscillator, described 
by an equation $\mathbf{\dot X=F(X)}$, where $\mathbf{X}$ is an 
$N$-dimensional, $N\ge2$, state vector.
Suppose that this system has a $T$-periodic limit 
cycle $\mathbf{X_T}(t+T)=\mathbf{X_T}(t)$. Then, for all $\mathbf{X}$ in 
the basin of attraction of $\mathbf{X_T}$, it is possible 
to introduce the phase $\vp(\mathbf{X})$ such that 
\[
\dot\vp(\mathbf{X})=\frac{2\pi}{T}=\omega\;.
\]
Essential for the definition of the phase is the notion of isochrons
\cite{Guckenheimer-75} as the sets of constant phase. 
These are the $(N-1)$-dimensional hypersurfaces $I_\vp$ such that  
$\vp(\mathbf{X})=\text{const}$ for $\mathbf{X}\in I_\vp$. Isochrons exist
in a basin of attraction of a stable limit cycle, but only in some exceptional cases
they can be expressed analytically. 

Consider now a coupled or driven system, described by 
$\mathbf{\dot X=F(X)}+\e\mathbf{\tilde p(X},t)$, where $\e$ quantifies 
the strength of coupling/driving. In this paper we will consider
the case of a periodic 
driving $\mathbf{\tilde p(X},t)=\mathbf{\tilde p(X},t+T_d)$. Then,
one can introduce the phase of the driving according to $\dot\psi=\nu=2\pi/T_d$ and write
the forcing term as a $2\pi$-periodic function of this phase $\mathbf{p(X},\psi)$.
To perform the phase reduction in the first approximation, one writes the equation for the 
phase $\vp(\mathbf{X})$:
\begin{equation}
\begin{aligned}
\dot\vp &=\frac{\partial \vp}{\partial \mathbf{X}}\mathbf{\dot X}=
\frac{\partial \vp}{\partial \mathbf{X}}[\mathbf{F(X)}+\e\mathbf{p(X},\psi)]=\\
&=\omega+\e\frac{\partial \vp}{\partial \mathbf{X}}\mathbf{p(X},\psi)\approx
\omega+\e\left.\frac{\partial \vp}{\partial \mathbf{X}}\right|_\mathbf{X_T}
\mathbf{ p(X_T},\psi)\;.
\end{aligned}
\label{eq:p1}
\end{equation}
Here in the last line one takes, in the first approximation in $\e$, the values of the
derivative of the phase and of the force on the limit cycle, where $\mathbf{X_T}=
\mathbf{X_T}(\vp)$. The resulting coupling term on the r.h.s. of \eqref{eq:p1}
is thus a function of the phases $\vp,\psi$:
\begin{equation}
Q_1(\vp,\psi)=
\left.\frac{\partial \vp}{\partial \mathbf{X}}\right|_\mathbf{X_T}\mathbf{ p(X_T},\psi)\;.
\label{eq:p2}
\end{equation}

We generalize this approach, representing the phase dynamics as an expansion
in powers of $\e$:
\begin{equation}
\dot \vp=\w+Q(\vp,\psi)=\w+\e Q_1(\vp,\psi)+\e^2 Q_2(\vp,\psi)+\e^3 Q_3(\vp,\psi)+\ldots\;.
\label{eq:series}
\end{equation}
Noteworthy, the adopted representation relies on the definition of 
the phase for the autonomous system, i.e. for $\e=0$; as mentioned above, 
an analytical relation between this phase and state variables $\mathbf{X}$ 
is generally unknown.
As we have seen, the existing theory provides only the linear in $\e$
 term $Q_1$ in 
Eq.~(\ref{eq:series}). Strictly speaking, the representation via 
a power series in $\e$ remains a conjecture - we will support it by the numerical 
analysis below. 

We now briefly discuss a special case when $\mathbf{p(X},\psi)=p(\psi)$,
i.e. the forcing term is a scalar (this means the force enters
only one equation in the system of ODEs for $\mathbf{X}$)
independent of the state of the system.
Then, according to Eq.~\eqref{eq:p2}, the first-order coupling function can be 
written as a product, $Q_1=Z(\vp)p(\psi)$, 
and the phase dynamics equation in the first approximation takes the so-called
Winfree form \cite{Winfree-80}:
\begin{equation}
\dot \vp=\w+\e Z(\vp)p(\psi)\;.
\label{eq:Win}
\end{equation}
The function $Z(\vp)=\left.\frac{\partial \vp}{\partial \mathbf{X}}\right|_\mathbf{X_T}$ 
is called phase sensitivity function or 
phase response curve (PRC).

A further reduction of the phase dynamics can be obtained if the
norm of the function $Q_1$ is small compared to $\w$. In this case the phase evolution
can be represented as a fast uniform rotation plus relatively slow additions.
This allows for averaging over the basic period, keeping only resonant terms
in the coupling function. The reason is that only such terms can
cause large, though slow, deviations of the phase from a uniform rotation. 
Which terms are resonant, 
depends on the relation between the autonomous frequency and the frequency of 
the forcing $\nu$. Namely, if $\w/\nu\approx m/n$, then the averaging yields the 
Kuramoto-Daido model~\cite{Kuramoto-84,Sakaguchi-Kuramoto-86,Daido-92a,Daido-93,Daido-96,Daido-96a}:
\begin{equation}
\dot \vp=\w+\e h(n\vp-m\psi)\;.
\label{eq:dai}
\end{equation}

\section{Phase reduction for the Stuart-Landau oscillator}
Our basic model is the forced Stuart-Landau oscillator (SLO)
\begin{equation}
\dot A=(\mu+i\eta)A-(1+i\alpha)|A|^2A+\e p(\psi)\;,
\end{equation}
where $A=Re^{i\theta}$ is the complex amplitude. This
equation is widely used a prototypic example of self-sustained 
oscillations, see, e.g., \cite{Nakagawa-Kuramoto-93,Nakagawa-Kuramoto-94,PhysRevLett.106.254101,%
Bordyugov-Pikovsky-Rosenblum-10,Sethia-Sen-14,Rosenblum-Pikovsky-15,Wilson-Ermentrout-18,%
Monga_Wilson-Matchen-Moehlis-18}.
The main advantage 
of this model is that the phase and the first-order
coupling function can be determined analytically, what simplifies the numerical
analysis of higher-order terms. 
It is convenient to re-write the model as a system
\begin{equation}
\begin{array}{rcl}
\dot{R}&=& \mu R-R^3 + \e p(\psi)\cdot\cos\theta\;,\\
\dot{\theta}&=& \eta -\alpha R^2-\e p(\psi)\cdot\sin\theta/R\;. 
\label{eq:polar}  
\end{array}
\end{equation}
Here $\alpha$ is the nonisochronicity parameter.  
For the autonomous oscillator, parameter $\mu$ determines the
radius $R_0=\sqrt{\mu}$ 
and stability of the limit cycle, while $\eta$, in combination with $\alpha,\mu$, 
determines the frequency of the oscillation. 

As is well-known (see, e.g.,~\cite{Pikovsky-Rosenblum-Kurths-01}), 
the phase of the autonomous SLO is defined as
\begin{equation}
\vp=\theta-\alpha \ln (R/R_0)\;.
\label{eq:phase}
\end{equation}
For the forced system, differentiating Eq.~(\ref{eq:phase}) with respect 
to time and substituting 
$\dot R$, $\dot\theta$ from Eq.~(\ref{eq:polar}), we obtain:
\begin{equation}
\dot\vp=\w -\frac{\alpha\cos\theta+\sin\theta}{R}\e p(\psi)\;,
\label{eq:num}
\end{equation}
where we introduced $\w=\eta-\alpha\mu$. If the forcing is so weak 
that the deviation from the 
limit cycle can be neglected, $R\approx R_0=\sqrt{\mu}$, then 
$\vp\approx \theta$ and Eq.~(\ref{eq:num}) yields the 
known first-order phase dynamics reduction for the SLO 
in the Winfree form, see Eq.~(\ref{eq:Win}), 
with the PRC
\begin{equation}
Z(\vp)=-\left(\alpha\cos\vp+\sin\vp\right)/\sqrt{\mu}\;.
\label{eq:phappr}
\end{equation}
For a harmonic forcing $p(\psi)=\cos\psi=\cos(\nu t )$ we obtain 
\begin{equation}
\begin{aligned}
Q_1&=-\left(\alpha\cos\vp+\sin\vp\right)\cos\psi/\sqrt{\mu}\\
&=-\frac{\alpha}{2\sqrt{\mu}}\left[\cos(\vp-\psi)+\cos(\vp+\psi)\right]-\frac{1}{2\sqrt{\mu}}
\left[\sin(\vp-\psi)+\sin(\vp+\psi)\right]\;.
\end{aligned}
\label{eq:Q1}
\end{equation}
Averaging $Q_1$ for $\nu\approx \w$ yields
\begin{equation}
h(\vp-\psi)=-\frac{\sin(\vp-\psi)}{2\sqrt{\mu}}-\frac{\alpha\cos(\vp-\psi)}{2\sqrt{\mu}}\;.
\label{eq:d1}
\end{equation}
As is well-known, this coupling function determines the synchronization domain of $1:1$ 
locking. Notice that other locked states do not appear in the averaged 
first-order approximation. 

\section{Computing nonlinear coupling function}
Here we present our numerical approach for determination 
of the nonlinear coupling function 
for the SLO.
We restrict ourselves, without loss of generality, 
to the case of harmonic driving $p(\psi)=\cos(\nu t)$, and proceed as 
follows. For some set of parameters $\mu,\alpha,\e,\nu$, 
we solve numerically Eqs.~(\ref{eq:polar}) and compute 
$\vp(t)$, $\dot\vp(t)$ with the help of Eqs.~(\ref{eq:phase},\ref{eq:num}).
Since the term $\w+\e Q_1(\vp,\psi)$ is known, we have to find only 
the nonlinear part of the coupling function
$Q_{nlin}=\e^2 Q_2(\vp,\psi)+ \e^3 Q_3(\vp,\psi)+\ldots$. 
For this purpose we fit the rest term 
$\dot\vp_r=\dot\vp-\w-\e Q_1(\vp,\psi)$ as a $2\pi$-periodic
function of variables $\vp,\psi$.
Practically, we perform a kernel-based estimation on  a grid
$100\times 100$, see \cite{Kralemann_et_al-13} for technical details. 
The error of the fit is quantified by 
\begin{equation}
\sigma=\text{std}[\dot\vp_r-Q_{nlin}(\vp,\psi)]/\text{std}[\dot\vp]\;,
\label{eq:err1}
\end{equation}
where $\text{std}[\xi]=\left [ \overline{(\xi-\bar{\xi})^2} \right ]^{1/2}$
and bar denotes the time averaging over the available time series.
The error $\sigma$ is due to a 
truncation of the series, to an error of the kernel estimator and to an error 
of the ODE solver. 
We emphasize that determination of $Q_{nlin}$ can fail for large $\e$ if, e.g., 
the SLO becomes entrained to the force. Indeed, in case of synchrony with the force, 
the trajectory does not cover the torus spanned by $\vp,\psi$, and the function 
of these two variables cannot be recovered. 
Generally, a strong force can also result in destruction of the smooth torus or make the 
torus so ``thick'' and shifted with respect to the original limit cycle that 
some loops cross one isochron twice, see a discussion in Ref.~\cite{Rosenblum-Pikovsky-19}.
In both latter cases the approach also fails.
This failure can be detected by monitoring the value of $\sigma$ which is for good cases
quite small.

The next task is to determine the basis functions $Q_k$ in the power series representation 
by Eq.~(\ref{eq:series}). 
For this goal we perform the above described computation of $Q_{nlin}$ for 
a fixed frequency $\nu=\text{const}$, and a set of values of the force amplitude $\e$
and then compute $Q_k$, $k\geq 2$, performing a polynomial fit in $\e$. 
(Recall that $Q_1$ is given by Eq.~(\ref{eq:Q1}).)
Practically, we truncate the series and
obtain only three terms $Q_{2,3,4}$ by fitting each 
element of $Q_{nlin}(\vp,\psi;\e)/\e^2$ by a second-order 
polynomial in $\e$, i.e. as $Q_{nlin}/\e^2\approx Q_2+\e Q_3+\e^2Q_4$.
The quality of this step is quantified by 
\begin{equation}
\gamma(\e)=\text{STD}\left[ Q_{nlin}-\e^2 Q_2-\e^3 Q_3-\e^4 Q_4\right]/\text{std}\left[Q\right]\;.
\label{eqsig123}
\end{equation}
Here $\text{STD}\left[\xi\right]=\langle(\xi-\langle \xi \rangle)^2\rangle^{1/2}$
and the averaging is performed as integration over the torus on which the coupling function
is defined:
\[
\langle w\rangle=(4\pi^2)^{-1}\int_0^{2\pi}{\rm d}\vp\int_0^{2\pi}{\rm d}\psi\, w(\vp,\psi)\;.
\]

\section{Nonlinear coupling functions for the SLO: results}
\subsection{Full nonlinear coupling function}
\begin{figure}[!h]
\centering\includegraphics[width=\textwidth]{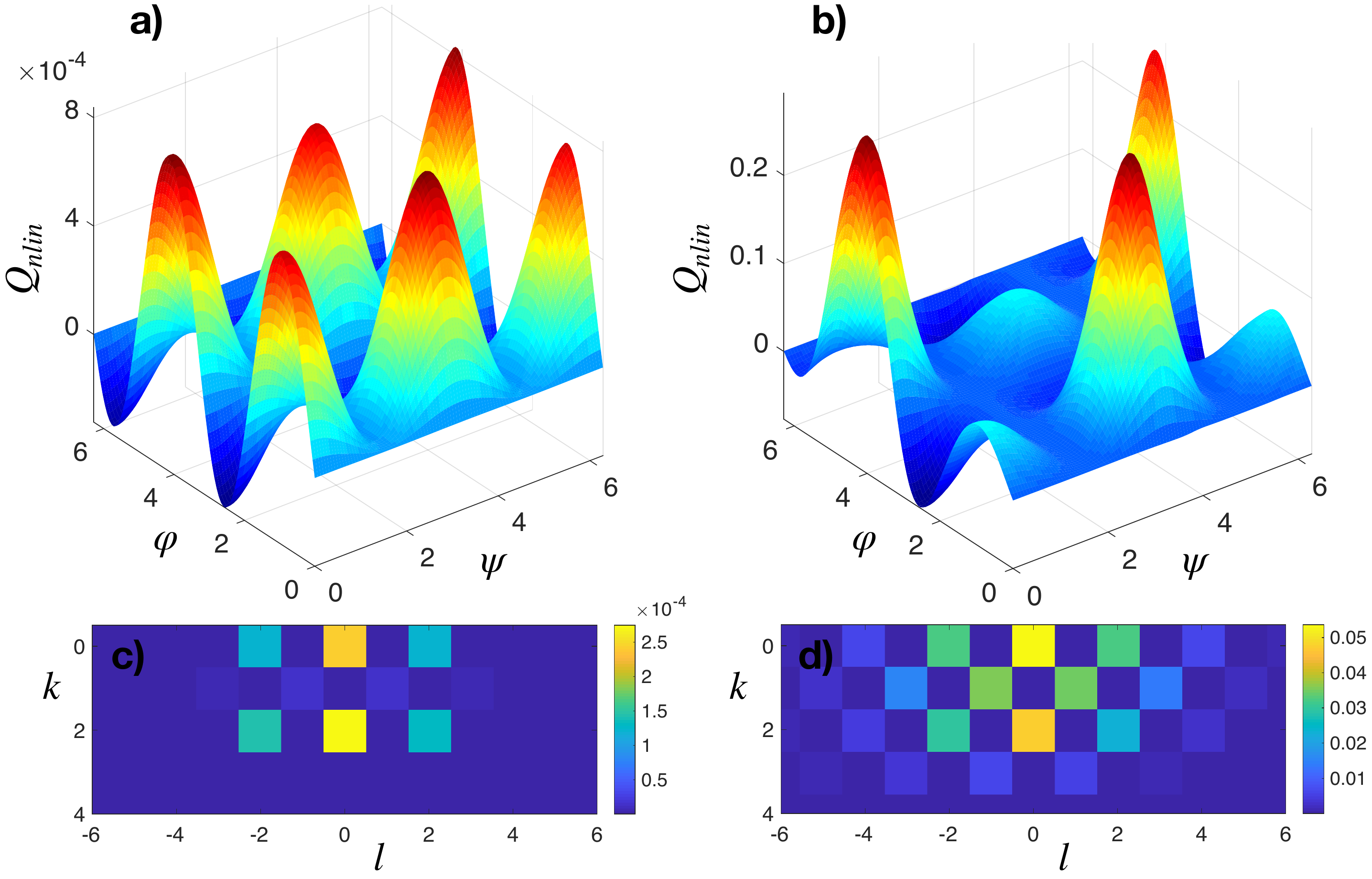}
\caption{Nonlinear part of the coupling function and the amplitudes of 
its Fourier modes (see~\eqref{eq:f}) for $\nu=0.3$ and 
$\varepsilon=0.05$ (panels (a,c)) and $\varepsilon=0.55$ (panels (b,d)).
}
\label{fig:Q03nlin}
\end{figure}

\begin{figure}[!h]
\centering\includegraphics[width=\textwidth]{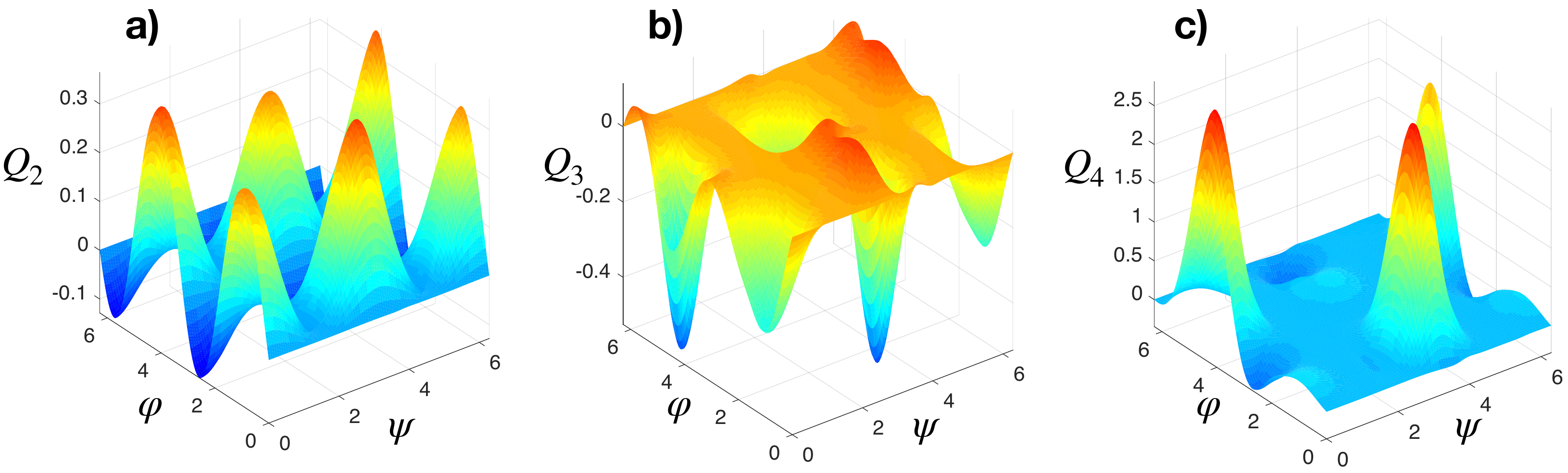}
\caption{The components $Q_{2,3,4}$ of the nonlinear coupling function for $\nu=0.3$,
obtained via a polynomial fit in the range $\e=0.05,\ldots,0.55$. 
Comparison with
Fig.~\ref{fig:Q03nlin} shows that $Q_2$ dominates, as expected, for small $\e$,
while $Q_4$ dominates for large $\e$.
}
\label{fig:Q03}
\end{figure}

In the first tests we compute the nonlinear coupling function
$Q_{nlin}$ and functions $Q_{2,3,4}$ for a fixed frequency of the force, 
$\nu=0.3$, and for different forcing amplitudes $\e$.  
Other parameters are $\eta=1$, $\mu=1$, $\alpha=0$, and we used $10^7$ data points 
for construction of $Q_{nlin}$. We obtained a good reconstruction for $\e\le0.55$: 
the error of the fit  $\sigma$, see Eq.~(\ref{eq:err1}), was smaller than $4\cdot10^{-3}$. 
For stronger forcing the system is close to synchronization with the force; here the 
reconstruction is poor and provides a non-smooth coupling function.  
The results are shown in Fig.~\ref{fig:Q03nlin}.
Here together with the shapes of $Q_{nlin}$ we show the amplitudes 
of Fourier modes of these functions,
defined according to
\begin{equation}
Q(\vp,\psi)=\sum_{k,l} F_{(k,l)} e^{ik\vp+il\psi}\;.
\label{eq:f}
\end{equation}
We remind, that the first-order coupling function
contains only harmonics $k=\pm 1,l=\pm 1$ (see~\eqref{eq:Q1}). 
One can see
that the shape of the nonlinear coupling function is very different 
from the linear one and depends strongly on $\e$. The components
$Q_{2,3,4}$ are illustrated in Fig.~\ref{fig:Q03}, all of them contain higher
Fourier modes. (The error of the power series representation is $\gamma(\e)<6.5\cdot10^{-3}$).

\begin{figure}[!h]
\centering\includegraphics[width=0.48\textwidth]{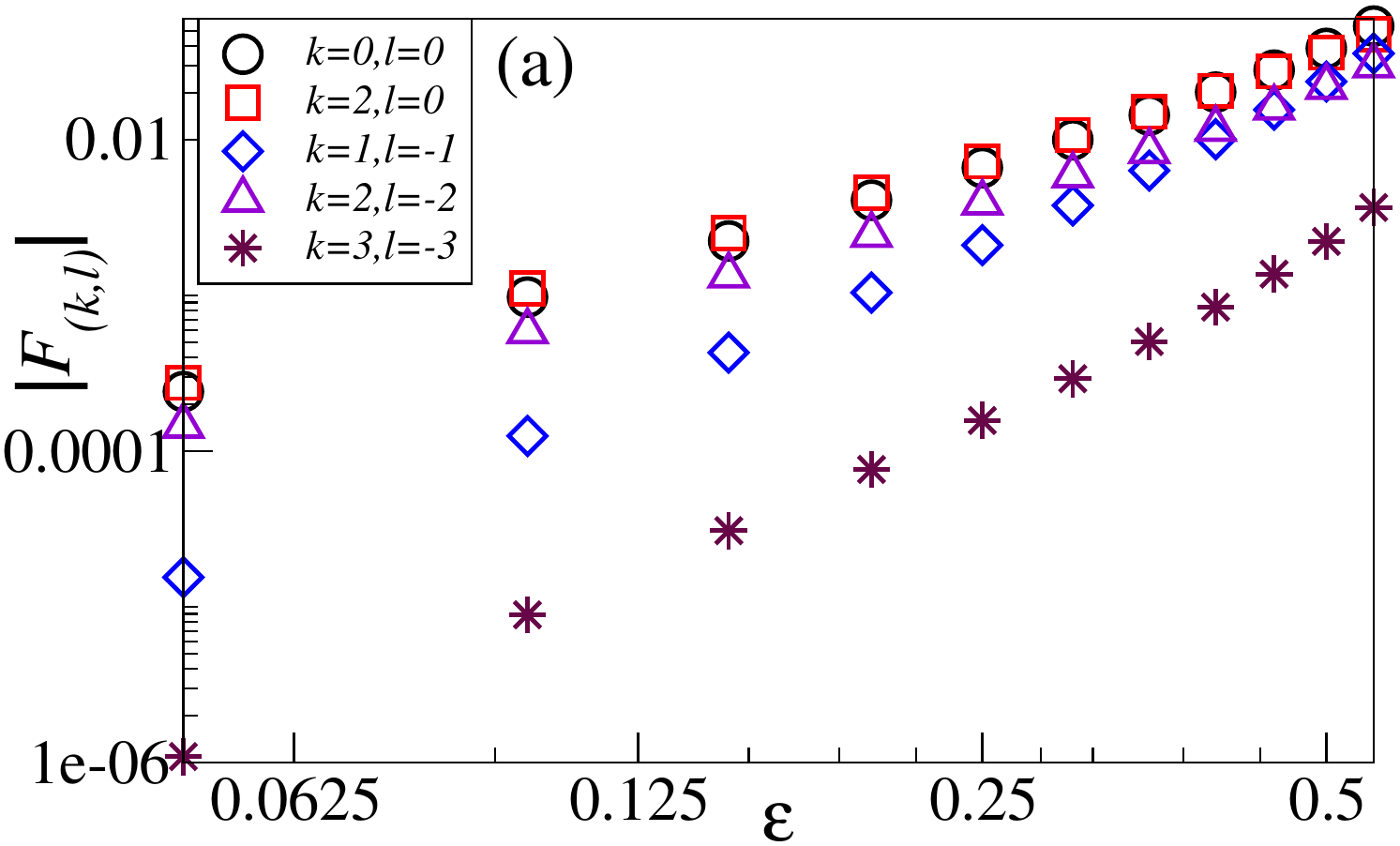}
\hfill\includegraphics[width=0.465\textwidth]{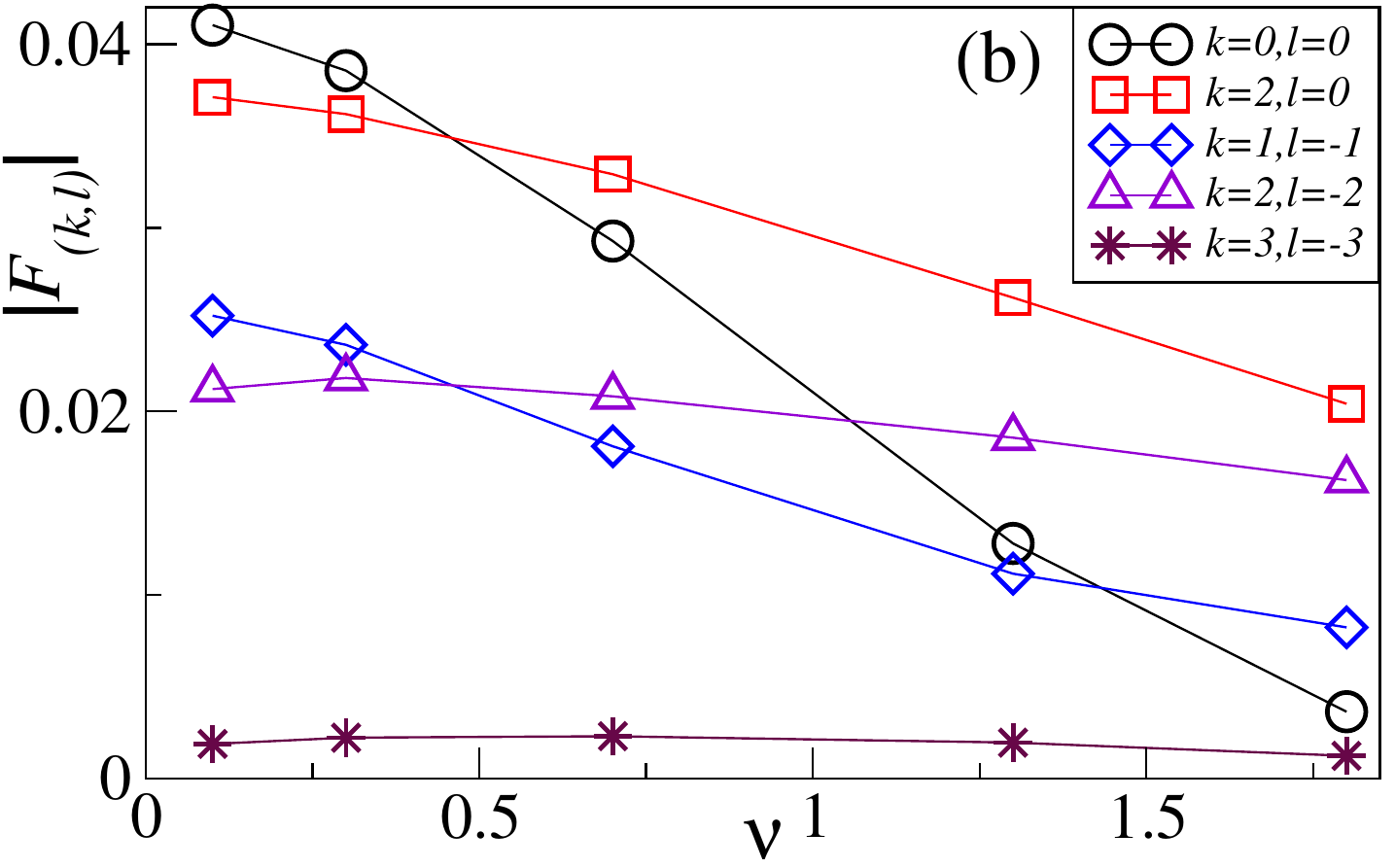}
\caption{Dependence of amplitudes of different Fourier modes on the 
amplitude $\e$ (a)
and on the frequency $\nu$ (b) of the forcing. }
\label{fig:Q03coef}
\end{figure}

As discussed above, the novel essential feature of the nonlinear coupling
function is its dependence on the frequency of the forcing $\nu$.
In Fig.~\ref{fig:Q03coef} we show dependencies of several dominant
Fourier modes of the coupling function on parameters $\e$ and $\nu$.

\begin{figure}[!h]
\centering\includegraphics[width=2.5in]{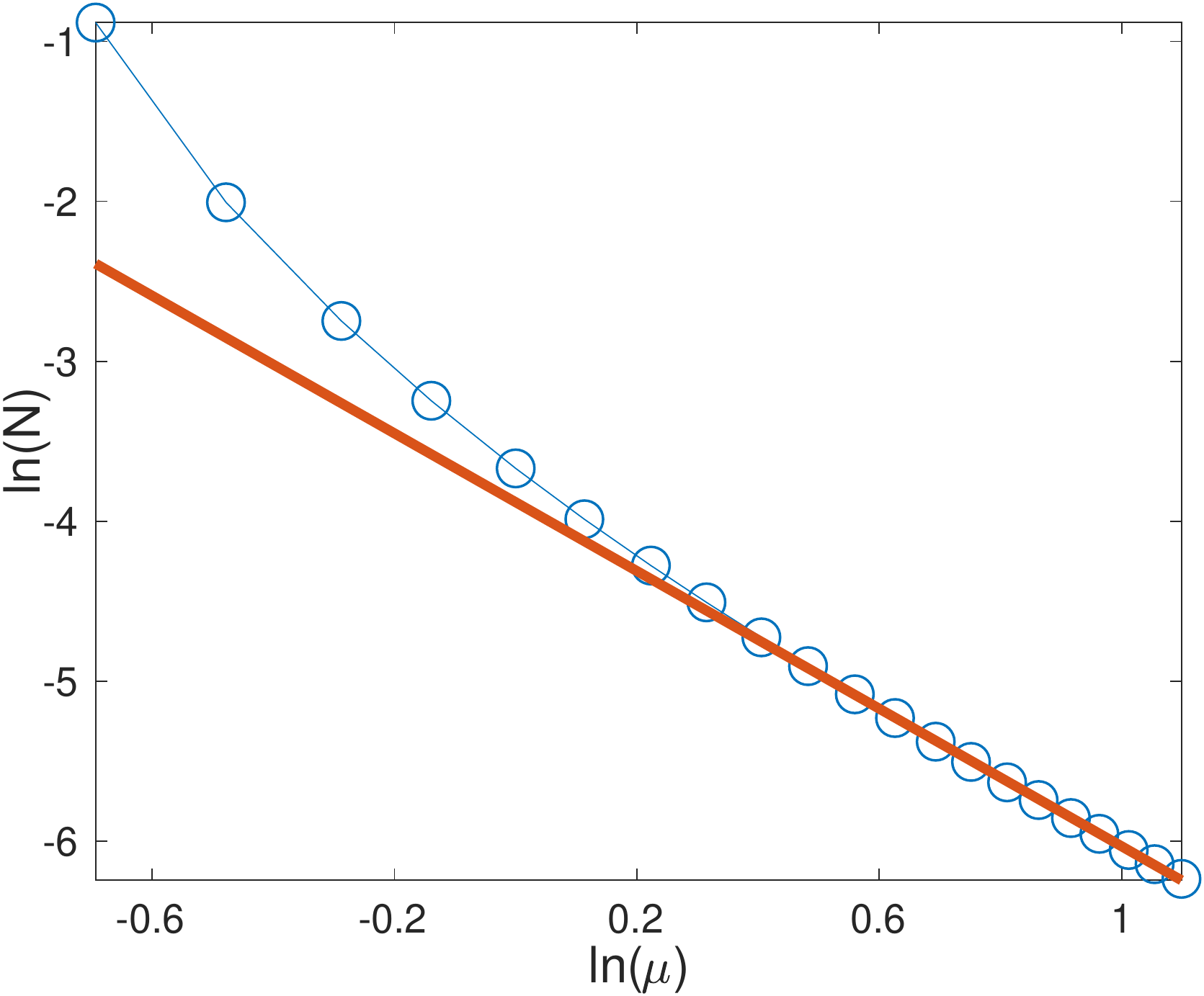}
\caption{The norm  of the nonlinear coupling function vs. parameter $\mu$ for constant
$\e$ (circles). 
Bold red line has slope $-2.15$.}
\label{fig:mudep}
\end{figure}

Next, we analyzed how the nonlinear coupling function varies with the parameter $\mu$.
As it follows from the first equation of (\ref{eq:polar}), this parameter determines the 
radius of the limit cycle oscillation $R_0=\sqrt{\mu}$. Furthermore, linearization of this 
equation yields for a small radius deviation $\delta$ from the limit cycle 
$\dot\delta\approx-2\mu\delta +\e p(t)\cos\theta$, so that the larger the value of $\mu$, 
the more stable is the cycle. We computed the nonlinear part of the coupling 
function for $0.5\le\mu\le 3$ and fixed parameters of the forcing, $\nu=0.1$, $\e=0.4$.
(For $\mu<0.5$ the forcing becomes too strong to provide a reliable 
construction of $Q_{nlin}$.)
The results are shown in Fig.~\ref{fig:mudep}. One can see that for large
$\mu$ the norm of the nonlinear coupling function decays as $\sim \mu^{-2.15}$,
what means that the nonlinear effects become less visible in the $\mu\to\infty$ limit,
because the linear part decays as $\sim \mu^{-1/2}$.

Finally, our simulations have shown no essentially interesting dependence
on parameter $\alpha\ne 0$, only some quantitative changes.

\subsection{Validity of the Winfree and the Kuramoto-Daido forms}
\begin{figure}[!h]
\centering\includegraphics[width=0.7\textwidth]{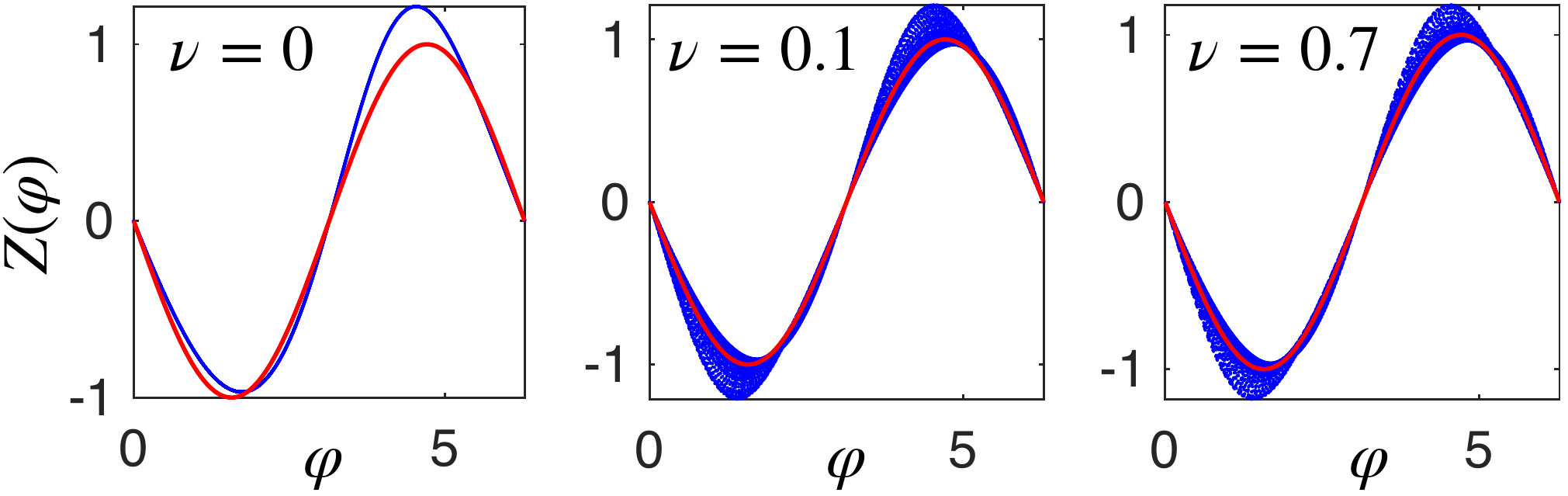}
\caption{Test for validity of the 
Winfree form of the phase coupling function, for $\e=0.4$ and different
values of frequency $\nu$.
Red lines show the first-order PRC according to Eq.~(\ref{eq:phappr}). 
}
\label{fig:PRCs}
\end{figure}

While the first-order coupling function for the forcing
term adopted in this study can be represented in the Winfree form,
this is no more valid for the full nonlinear coupling function.
 In order to check the validity of the Winfree 
representation for strong forcing, we estimate an
``effective'' $Z(\vp)$ by 
plotting $\frac{\dot\vp-\w}{\e\cos(\nu t)}$ vs $\vp$ 
for $\e\cos(\nu t)>10^{-5}$, cf. Eq.~(\ref{eq:Win}). 
The results for $\e=0.4$ and three different values of $\nu$ 
are presented in Fig.~\ref{fig:PRCs}. For a constant perturbation, $\nu=0$,
this approach yields a curve that, as expected, deviates from the 
linear PRC given by Eq.~(\ref{eq:phappr}). However, for harmonic driving, 
the points in the plot do not fall on a curve, what means that in the nonlinear regime 
the coupling function cannot be decomposed into a product $Z(\vp)p(\psi)$.

One could find an approximate PRC by averaging the curves in Fig.~\ref{fig:PRCs},
or by neglecting all the Fourier-components in the expansion~\eqref{eq:f} 
except for $l=1$ (and taking only real part of it). In this way one, however,
neglects terms that are of the same order of magnitude as the preserved ones. 

\begin{figure}[!h]
\centering\includegraphics[width=4in]{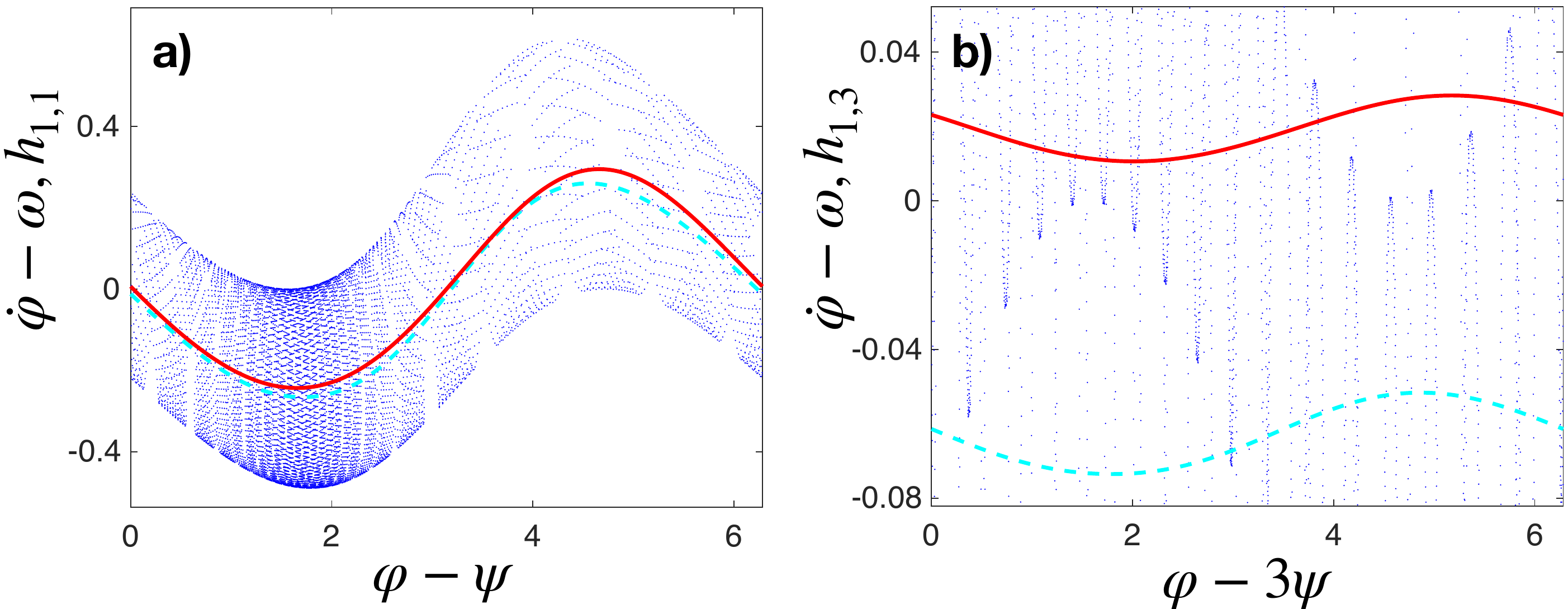}
\caption{The Kuramoto-Daido models $h(\varphi-\psi)$ constructed  for $\e=0.5$ 
and $\nu=0.7$ (a) and $h(\varphi-3\psi)$ for $\e=0.5$, $\nu=0.3$. 
Blue points show exact values of phase derivative. Solid red and dashed cyan curves are
the results of two ways of averaging: 
the first one is obtained by taking the corresponding diagonal Fourier modes of
the reconstructed full coupling function $Q(\vp,\psi)$,
while the second one is obtained via a direct Fourier fit of blue dots.
}
\label{fig:D}
\end{figure}

As was discussed in Section~\ref{sec2}, there is no unique Kuramoto-Daido model,
rather there is a set of models valid for different resonances $\w/\nu\approx m/n$.
The coupling function for the resonance $\w/\nu\approx m/n$ is obtained from
the full coupling function~\eqref{eq:f} as
\[
h_{n,m}(n\vp-m\psi)=\sum_{k}F_{(kn,-km)}\exp[ikn\vp-ikm\psi]\;.
\]
For example, the main resonance Kuramoto-Daido coupling function $h_{1,1}$ is described by
the harmonics $F_{(0,0)},F_{(\pm 1,\mp 1)}$, $F_{(\pm 2,\mp 2)},F_{(\pm 3,\mp 3)},\ldots$
 In the first-order approximation one has just the first harmonics terms~\eqref{eq:d1}, while
for the full nonlinear coupling function also higher-order terms are present for the main
resonance. For other resonances, which are not present in the first order,
nonlinear coupling provides effective averaged resonant forcing in higher orders
in $\e$. Another way
to construct the Kuramoto-Daido model is to perform a direct 
 fit of $\dot\vp-\w$ vs
$(n\vp-m\psi)\,\text{mod}\,2\pi$ (e.g., representing the function as a Fourier series and finding
the Fourier coefficients through minimization of the mean squared error),
this approach have been adopted in Ref.~\cite{Tokuda-Jain-Kiss-Hudson-07}.
 We illustrate the Kuramoto-Daido 
coupling functions $h_{1,1}$ and $h_{1,3}$ for $\e=0.5$
in Fig.~\ref{fig:D}. While $h_{1,1}$ is rather close to the first-order
Kuramoto-Daido model~\eqref{eq:d1}, the norm of the coupling for the resonance $1:3$
is rather small.

\section{Predicting synchronization regions with nonlinear coupling functions}
In this section we demonstrate that the
nonlinear phase model can be exploited to predict 
locking regions, or Arnold tongues.
We recall that we cannot construct the coupling function if the system is locked to 
an external force. However, it does not mean that the phase model is not valid in that 
parameter domain, but simply that our procedure for the coupling function construction 
fails. Nevertheless, we can use the coupling function obtained for coupling strength below
the synchronization threshold to predict domain of synchrony for stronger forcing (or for
other frequencies of the forcing). 

\begin{figure}[!h]
\centering\includegraphics[width=2.5in]{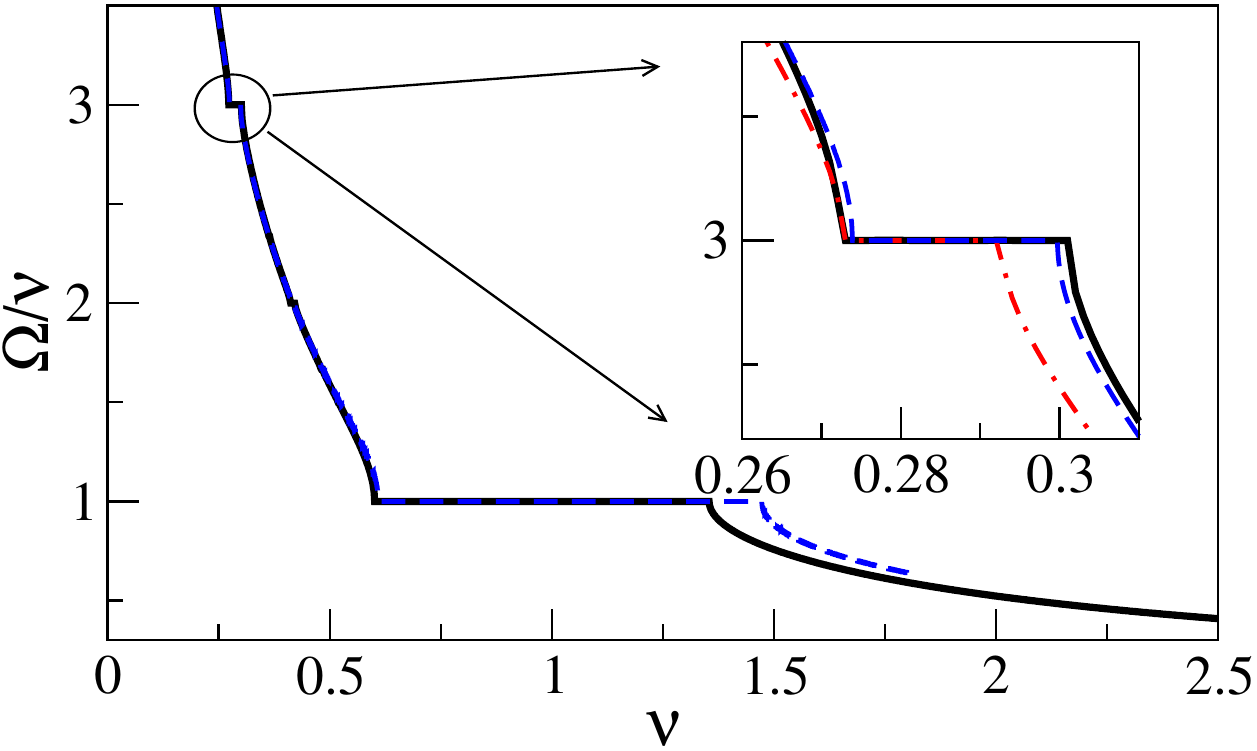}
\caption{Synchronization domains of the forced
SLO with $\e=0.7$. Black solid line shows the true devil's staircase and
the blue dashed one shows the prediction by the full nonlinear 
phase model. The red dashed-dotted line is obtained from the integration of the first-order
phase approximation~\eqref{eq:Q1}; one can see that it provides a significantly worse
prediction compared to the full model.
}
\label{fig:synchro}
\end{figure}

For small
amplitudes of the forcing, only the main Arnold tongue with $\nu\approx \omega$
is relevant, and it is well captured by
the Kuramoto-Daido representation of the phase dynamics in terms 
of phase differences, cf. Eq.~(\ref{eq:d1}). 
This form of coupling determines the only synchronization domain that has a 
triangular shape and touches the $\nu$-axis.
In the strongly forced regime we can expect appearance of further Arnold tongues.
Indeed, the devil's staircase computed for the full model (\ref{eq:polar}) for 
$\e=0.7$,  exhibits not only $1:1$ locking but also domains of $1:3$ and $1:2$
synchrony, see Fig.~\ref{fig:synchro}. 

Now we check how this staircase
can be reproduced by the phase model, constructed for $\nu=0.3$, $\e\le 0.55$.
(We remind that for $\e>0.55$ the model construction failed because of synchrony.)
Combining Eq.~(\ref{eq:series}) with $\dot\psi=\nu$ we obtain
\begin{equation}
\frac{\rm d\vp}{\rm d\psi}=\left(\w +\e Q_1+\e^2Q_2+\e^3Q_3+\e^4Q_4\right )/\nu\;.
\end{equation} 
Next, we solve this equation numerically for $\e=0.7$. 
Namely, using the Euler technique and precomputed $Q_{2,3,4}$, 
we find phase increase $\Delta\vp$ 
corresponding to a large
phase increase $\Delta\psi$ and obtain frequency ratio as $\Delta\vp/\Delta\psi$.
The result is shown in Fig.~\ref{fig:synchro}. We see that the phase model obtained 
for $\nu=0.3$ very well describes the $1:3$ locking domain and the left border of
the $1:1$ locking region, but exhibits an essential deviation at the right border
of the latter. This can be explained by the frequency dependence of $Q_{nlin}$.

\begin{figure}[!h]
\centering\includegraphics[width=2in]{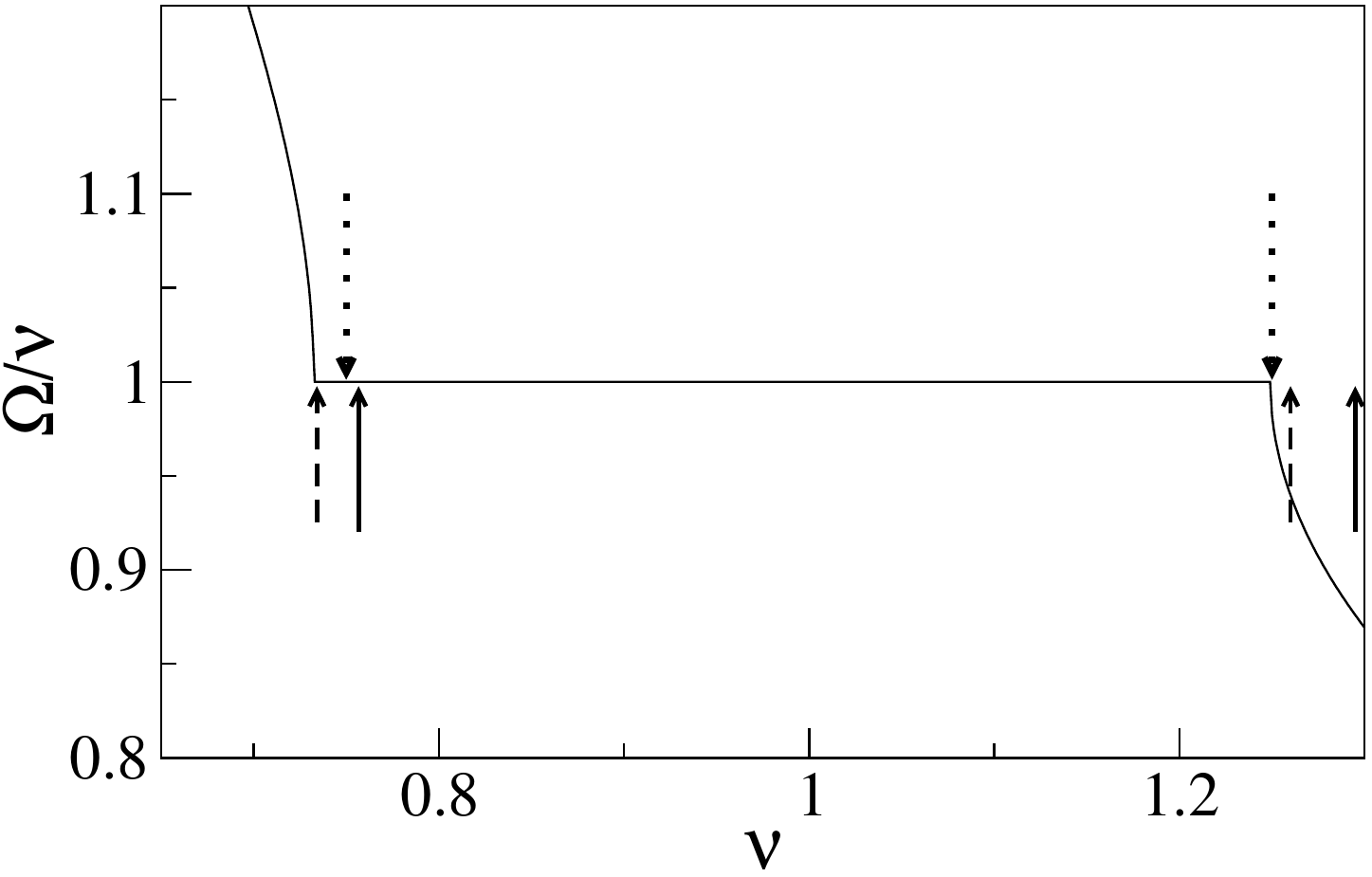}
\caption{The  $1:1$ synchronization domain for the SLO forced with
amplitude $\e=0.5$ and its 
borders predicted by the 
Kuramoto-Daido model reconstructed for $\nu=0.7$. 
The solid curve shows the true locking region.
The solid lines with arrows show the borders predicted by the the model 
obtained from the Fourier coefficients of the coupling function; the 
dashed lines with arrows show the corresponding prediction of 
model obtained via direct Fourier fit.
}
\label{fig:tonD}
\end{figure}

As has been discussed above, the Kuramoto-Daido model is expected to
be good for small forcing only, because for large forcing the time
scale separation between the uniform phase rotation and deviations from it 
is not valid. Nevertheless, one can formally apply
this model and to check quality of  predictions
of synchronization properties for large forcing amplitudes.  We illustrate,
how good the model based on the coupling
function $h_{1,1}$ predicts the boundaries of the main synchronization
region $1:1$ in Fig.~\ref{fig:tonD}. One can see that the prediction
is quite reasonable, what indicates that for the synchronization properties
many nonlinear features of the coupling function are not important.
While the Kuramoto-Daido coupling function $h_{1,3}$ (see Fig.~\ref{fig:D}(b))
correctly predicts existence of the synchronization region $1:3$,
its position is strongly shifted in $\nu$ compared to the really
observed one, therefore we do not depict
it in Fig.~\ref{fig:synchro}.

\vspace*{-5pt}

\section{Conclusion}
In summary, we have presented the concept of a nonlinear phase
coupling function for a periodically forced self-sustained oscillator.
It generalizes the approach of the phase reduction based on the first order
approximation in the forcing strength. For illustration we have chosen
the Stuart-Landau oscillator, mainly for the reason
of convenience of presentation, because for it the phase and 
the first-order phase reduction are known analytically. The method
can be however straightforwardly applied to other systems, for which the
dynamical equations are known. In such a case, 
the proper phase and its derivative should be determined numerically, 
see \cite{Rosenblum-Pikovsky-19}.
The case of a purely observational determination of the nonlinear coupling 
function (cf.~\cite{Kralemann_et_al-13,RevModPhys.89.045001}) 
requires additional efforts, as the 
reliable methods of the proper phase
reconstruction from scalar signals are still missing.

We have demonstrated that the nonlinear coupling function has a shape
quite different from that of the first-order approximation, with many
more Fourier components present. A novel feature is a dependence of the
nonlinear terms on the frequency of the forcing, in contradistinction
to the first approximation which is frequency-independent. We have also
shown that many differences between the full nonlinear coupling function 
and its first-order approximation are not
so important for determination of the synchronization regions, although the
full nonlinear function provides better accuracy.

We foresee that the presented approach can be extended to determination
of the phase dynamics of coupled oscillators at strong coupling. 
An extra problem to be treated here is an additional dependence of the forcing 
waveform on the strength of the coupling. 
This study will be reported elsewhere. 
\enlargethispage{20pt}

The study was supported by the Russian Science
Foundation (Grant No. 17-12-01534).  We thank S. Schaefer and E. Gengel for useful discussions.




\end{document}